\documentclass[12pt]{article}

\usepackage{epsfig}

\usepackage{amssymb}
\usepackage{amsmath}
\usepackage{amsfonts}

\def\be{\begin{equation}}
\def\ee{\end{equation}}
\def\ba{\begin{array}{c}}
\def\ea{\end{array}}

\def\ben{$$}
\def\een{$$}

\newcommand{\bea}{\begin{eqnarray}}
\newcommand{\eea}{\end{eqnarray}}

\newcommand{\kt}{\rangle}
\newcommand{\br}{\langle}
\begin{document}

\titlepage

\vspace{.35cm}

 \begin{center}{\Large \bf

Cryptohermitian Hamiltonians on graphs

  }\end{center}

\vspace{10mm}

 \begin{center}

 {\bf Miloslav Znojil}

 \vspace{3mm}
Nuclear Physics Institute ASCR,

250 68 \v{R}e\v{z}, Czech Republic

{e-mail: znojil@ujf.cas.cz}

\vspace{3mm}

\end{center}

\vspace{5mm}

\section*{Abstract}

A family of nonhermitian quantum graphs is proposed and studied via
their discretization.


\section{Why? Short-range nonlocalities}

The abstract Quantum Theory \cite{Messiah} admits that the physical
inner product of two wave functions defined, say, in the form of
integral
 \ben
 \br \phi | \psi \kt \sim \int dx\, \int dy \,\phi^*(x)
\Theta(x,y) \psi(y)
 \een
may be assumed slightly nonlocal\footnote{For the sake of
simplicity, we shall only consider the one-dimensional motion
here.}. In our recent paper \cite{fund} we discussed such an option
and assumed that the most popular ``Dirac-type" choice of
$\Theta(x,y)=\delta(x-y)$ is being ``smeared" over some microscopic
spatial domain. We paid particular attention to the use of kernels
$\Theta(x,y)$ which do not vanish for $|x-y|<\theta$ where the
``fundamental length scale" $\theta>0$ characterizes the
hypothetical smearing and nonlocality of the quantum model in
question.

%
\begin{figure}[h]                     
\begin{center}                         
\epsfig{file=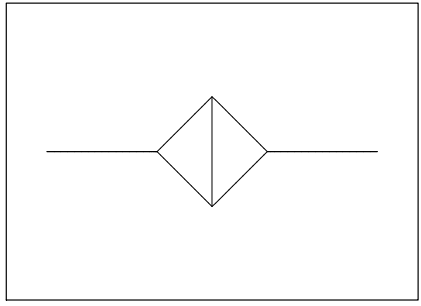,angle=270,width=0.4\textwidth}
\end{center}                         
\vspace{-2mm} \caption{A typical small-non-locality-simulating
graph.
 \label{fied}}
\end{figure}

An immediate practical consequence of the introduction of the
spatial smearing $\sim {\cal O}(\theta)$ is that the most common
(viz., Euclidian) representation of the 1D space by real line
$\mathbb{R}$ may be deformed and locally modified. On the purely
kinematical level every sufficiently short spatial interval of the
coordinate $x \in (a,a+\theta)$ may be replaced by its topologically
nontrivial graph-shaped alternative (cf., e.g., the example of
resulting toy-model set $\mathbb{G}$ of coordinates $x$ in
Fig.~\ref{fied}).

In our present paper we intend to study this possibility in more
detail.

\section{How? Discretizations}

\subsection{Topologically nontrivial models}

The first non-tree lattice or discrete graph will be selected here
in the form
 \be
 \ba
   \mbox{\ \ \ \ \ \ \ \
  }_\diagup\begin{array}{|c|}
 \hline
 {x_{0^+}}\\
 \hline
 \ea_\diagdown \\
  \ \begin{array}{||c||}
 \hline
 \hline
 {x_{-K}}\\
 \hline
 \hline
 \ea-\
 \ldots\   -\begin{array}{|c|}
 \hline
 {x_{-2}}\\
 \hline
 \ea-
 \begin{array}{||c||}
 \hline
 \hline
 \mbox{} x_{-1}\\
 \hline
 \hline
 \ea_\diagdown^\diagup
 \
 \begin{array}{c}
  \mbox{\ \ \
  \ \ \ \
 \ \ }\\
  \ea
 \ ^\diagdown_\diagup\begin{array}{||c||}
 \hline
 \hline
 \mbox{} x_1\\
 \hline
 \hline
 \ea - \begin{array}{|c|}
 \hline
 {x_{2}}\\
 \hline
 \ea -  \ldots -
   \begin{array}{||c||}
 \hline
 \hline
 {x_{K}}\\
 \hline
 \hline
 \ea  \\
   \mbox{\ \ \ \ \ \ \ \
 }^\diagdown\begin{array}{|c|}
 \hline
 {x_{0^-}}\\
 \hline
 \ea^\diagup  \\
  \ea
  \,
 \label{VoooKL}
 \ee
possessing four vertices $x_{-K}$,  $x_{-1}$,  $x_{1}$ and $x_{K}$
and four wedges. For the sake of simplicity, just the external
wedges will be of variable length, $K=1,2,\ldots$.

We shall preserve the most common form of the discrete Laplacean
$\triangle$ on this lattice, with a variable weight $ u$ in
 \ben
 \triangle \psi(\xi_{k}) \sim  -
 \frac{\psi(\xi_{k+1})-u\,\psi(\xi_{k})+\psi(\xi_{k-1})}{h^2}\,,
 \ \ \ \ \ k \neq \pm 1
  \een
as well as in
 \ben
 \triangle \psi(\xi_{k}) \sim  -
 \frac{\psi(\xi_{j})-u\,\psi(\xi_{k})+\psi(\xi_{0^+})+\psi(\xi_{0^-})}{h^2}\,,
 \ \ \ \ \ \ j=2k\,,
 \ \ \ \ k=\pm 1\,.
  \een
Although the  choice of the weights $u$ is, in principle, amenable
to variations, its present assignment to individual grid points will
be controlled by the following allocations,
 \ben
 \ba
   \mbox{\ \
  }_\diagup\begin{array}{|c|}
 \hline
 {2}\\
 \hline
 \ea_\diagdown \\
  \begin{array}{||c||}
 \hline
 \hline
 {2}\\
 \hline
 \hline
 \ea-\
 \ldots\   -\begin{array}{|c|}
 \hline
 {2}\\
 \hline
 \ea-
 \begin{array}{||c||}
 \hline
 \hline
 \mbox{} 3\\
 \hline
 \hline
 \ea_\diagdown^\diagup
 \
 \begin{array}{c}
  \mbox{\ \ \
  \
 \ \ }\\
  \ea
 \ ^\diagdown_\diagup\begin{array}{||c||}
 \hline
 \hline
 \mbox{} 3\\
 \hline
 \hline
 \ea - \begin{array}{|c|}
 \hline
 {{2}}\\
 \hline
 \ea -  \ldots -
   \begin{array}{||c||}
 \hline
 \hline
 {2}\\
 \hline
 \hline
 \ea  \\
   \mbox{\ \
 }^\diagdown\begin{array}{|c|}
 \hline
 {2}\\
 \hline
 \ea^\diagup  \\
  \ea
  \,.
 \label{VooindoKL}
 \een
Naturally, we could keep the related Hamiltonian purely kinetic. In
such a case \cite{web}, the detailed form of the Hamiltonian matrix
is to be derived from the assumption that all the points of the
lattice are ordered in the sequence $x_{-K}$, $\ldots$, $x_{-2}$,
$x_{-1}$, $x_{0^-}$, $x_{0^+}$, $x_{1}$, $x_{2}$, $\ldots$, $x_{K}$.
The presence of the loop in lattice (\ref{VoooKL}) only implies that
the coordinate subscript has to run over the straightened sequence
${-K}$, $\ldots$, ${-2}$, ${-1}$, ${0^-}$, ${0^+}$, ${1}$, ${2}$,
$\ldots$, ${K}$. In this way we arrive at the matrix Hamiltonian
 \ben
 H=
 \left[ \begin {array}{cccccccccccc}
  2&-1&&&&&&&&&&\\
 - 1&2&\ddots&&&&&&&&&\\
  &\ddots&\ddots&-1&&&&&&&&\\
 &&- 1&2&-1&&&&&&&\\
 &&&-1&3&-1&-1&&&
 &&\\
  &&&&-1&2&&-1&&&&\\
 &&&&-1&&2&-1&&
 &&\\
 &&&&&-1&-1&3&-1&&&\\
 &&&& &&&-1&2&-1&&\\
 &&&&&&& &-1&\ddots&\ddots&\\
 &&&&&&&& &\ddots&2&-1\\
 &&&&&&&&&&-1&2
 \end {array} \right]\,.
 \een
Due to the topologically nontrivial origin of this matrix, only its
central partition deviates from the tridiagonal pattern.

We should add that in the same spirit a systematic refinement can be
considered for the loop in the center of our discrete graph. In this
manner one obtains the four-point loop and lattice
 \be
 \ba
   \mbox{\ \ \ \ \ \ \ \
  }_\diagup\begin{array}{|c|}
 \hline
 {x_{U_-}}\\
 \hline
 \ea-\begin{array}{|c|}
 \hline
 {x_{U_+}}\\
 \hline
 \ea_\diagdown \\
 \ \ \,
  \begin{array}{||c||}
 \hline
 \hline
 {x_{-K}}\\
 \hline
 \hline
 \ea   - \ldots -
 \begin{array}{|c|}
 \hline
 {x_{-2}}\\
 \hline
 \ea-
 \begin{array}{||c||}
 \hline
 \hline
 \mbox{} x_{-1}\\
 \hline
 \hline
 \ea_\diagdown^\diagup
 \
 \begin{array}{c}
  \mbox{\ \ \
  \ \ \ \
  \ \
  \ \ \ \
  \ \
  \ \ \ \
 \ \ }\\
  \ea
 \ ^\diagdown_\diagup\begin{array}{||c||}
 \hline
 \hline
 \mbox{} x_1\\
 \hline
 \hline
 \ea - \begin{array}{|c|}
 \hline
 {x_{2}}\\
 \hline
 \ea - \ldots -
   \begin{array}{||c||}
 \hline
 \hline
 {x_{K}}\\
 \hline
 \hline
 \ea    \\
   \mbox{\ \ \ \ \ \ \ \
 }^\diagdown\begin{array}{|c|}
 \hline
 {x_{D_-}}\\
 \hline
 \ea-\begin{array}{|c|}
 \hline
 {x_{D_+}}\\
 \hline
 \ea^\diagup  \\
  \ea
  \
 \label{V10KL3}
 \ee
or, in general, the $2L-$point circular sublattice representing the
loop.

\subsection{Cryptohermitian interactions \label{probe} }

In a parallel to the quantum-graph constructions of paper
\cite{fundgra} we shall endow the two central vertices $x_{-1}$ and
$x_{1}$ with a nontrivial interaction. In an expectation of
conversion of this interaction into a source of a nontrivial
fundamental length as described in Ref.~\cite{fund} we shall admit
that this interaction violates the Hermiticity of our Hamiltonian
matrix. This being said, the simplest illustrative example of the
resulting non-Hermitian Hamiltonian $H=H^{(K)}(g,h)$ assigned to
graph (\ref{VoooKL}), say, at $K=3$  will read
 \ben
 H^{(3)}(g,h)=
 \left[ \begin {array}{cccccccc} 2&-1&&&&&&\\
 -
 1&2&-1&&&&&\\
 &-1&3&-1-g&-1-h&&&
 \\
  &&-1+g&2&&-1+h&&\\&&-1+h&&2&-1+g
 &&\\
 &&&-1-h&-1-g&3&-1&\\
 &&&& &-1&2&-1\\
 &&&&&&-1&2
 \end {array} \right]\,.
 \een
A further natural generalization of the model will be obtained when
we append the same elementary Hermiticity-violating nearest-neighbor
interaction terms to the two outmost vertices $x_{-K}$ and $x_{K}$.
At $K=3$ this will lead to the three-parametric Hamiltonian
$H=H^{(K=3)}(g,h;z)$ acquiring the eight-dimensional sparse-matrix
form
%
 \ben
  \left[ \begin {array}{cccccccc} 2&-1-z&&&&&&\\
 -
 1+z&2&-1&&&&&\\
 &-1&3&-1-g&-1-h&&&
 \\
  &&-1+g&2&&-1+h&&\\&&-1+h&&2&-1+g
 &&\\
 &&&-1-h&-1-g&3&-1&\\
 &&&& &-1&2&-1+z\\
 &&&&&&-1-z&2
 \end {array} \right]\,.
 \een
Next, the  algebraization related, say, to graph (\ref{V10KL3}) with
$K=3$ and $L=2$,
 \be
 \ba
   \mbox{\ \ \ \ \ \ \ \
  }_\diagup\begin{array}{|c|}
 \hline
 {x_{U_-}}\\
 \hline
 \ea-\begin{array}{|c|}
 \hline
 {x_{U_+}}\\
 \hline
 \ea_\diagdown \\
 \ \ \,
  \begin{array}{||c||}
 \hline
 \hline
 {x_{-3}}\\
 \hline
 \hline
 \ea   -
 \begin{array}{|c|}
 \hline
 {x_{-2}}\\
 \hline
 \ea-
 \begin{array}{||c||}
 \hline
 \hline
 \mbox{} x_{-1}\\
 \hline
 \hline
 \ea_\diagdown^\diagup
 \
 \begin{array}{c}
  \mbox{\ \ \
  \ \ \ \
  \ \
  \ \ \ \
  \ \
  \ \ \ \
 \ \ }\\
  \ea
 \ ^\diagdown_\diagup\begin{array}{||c||}
 \hline
 \hline
 \mbox{} x_1\\
 \hline
 \hline
 \ea - \begin{array}{|c|}
 \hline
 {x_{2}}\\
 \hline
 \ea -
   \begin{array}{||c||}
 \hline
 \hline
 {x_{3}}\\
 \hline
 \hline
 \ea    \\
   \mbox{\ \ \ \ \ \ \ \
 }^\diagdown\begin{array}{|c|}
 \hline
 {x_{D_-}}\\
 \hline
 \ea-\begin{array}{|c|}
 \hline
 {x_{D_+}}\\
 \hline
 \ea^\diagup  \\
  \ea
  \
 \label{V10KL3b}
 \ee
will lead to the $N$ by $N$ matrix Hamiltonian $H^{(K,L)}(g,h;z)$
with dimension $N=2K+2L=10$, viz., to the matrix
 \ben
  \left[ \begin {array}{cccccccccc}
   2&-1-z&{}&{}&{}&{}&{}&{}&{}&{}
 \\
 -1+z&2&-1&{}&{}&{}&{}&{}&{}&{}\\
 {}&-1&3&
 -1-g&{}&-1-h&{}&{}&{}&{}\\
 {}&{}&-1+g&2&-1&{}&{}&{}&{}&{}
 \\
 {}&{}&{}&-1&2&{}&{}&-1+h&{}&{}\\
 {}&{}&-1+
 h&{}&{}&2&-1&{}&{}&{}\\
 {}&{}&{}&{}&{}&-1&2&-1+g&{}&{}
 \\
 {}&{}&{}&{}&-1-h&{}&-1-g&3&-1&{}\\
 {}&{}&{}
 &{}&{}&{}&{}&-1&2&-1+z\\
 {}&{}&{}&{}&{}&{}&{}&{}&-1-z&2
 \end {array} \right]\,
 \een
etc.

\section{Energies: Factorized secular equations \label{treti}}

The spectra of our toy-model Hamiltonians must be calculated
numerically in general. It is still instructive to keep in mind that
these spectra remain obtainable in closed form at the first few
integers~$K$.

\subsection{$K=1$: Loop-shaped discrete lattice }

On the degenerate single-loop $K=1$ lattice
 \be
 \ba
   \mbox{\ \ \ \ \ \ \ \
  }_\diagup\begin{array}{|c|}
 \hline
 {x_{0^+}}\\
 \hline
 \ea_\diagdown \\
 \ \ \ \ \ \
 \begin{array}{||c||}
 \hline
 \hline
 \mbox{} x_{-1}\\
 \hline
 \hline
 \ea_\diagdown^\diagup
 \
 \begin{array}{c}
  \mbox{\ \ \
  \ \ \ \
 \ \ }\\
  \ea
 \ ^\diagdown_\diagup\begin{array}{||c||}
 \hline
 \hline
 \mbox{} x_1\\
 \hline
 \hline
 \ea  \\
   \mbox{\ \ \ \ \ \ \ \
 }^\diagdown\begin{array}{|c|}
 \hline
 {x_{0^-}}\\
 \hline
 \ea^\diagup  \\
  \ea
  \,
 \label{Vooo4KL1}
 \ee
we may consider the most elementary Hamiltonian $H^{(1)}(g,h;z)$
with spectrum shown in Figure \ref{fionej2}. This picture clearly
shows that once we reparametrize $g=g(\gamma,\delta)=\gamma+\delta$
and $h=h(\gamma,\delta)=\gamma-\delta$, the factorization of the
secular equation is achieved. This empirical fact may be clarified
by elementary algebra which gives the quadruplet of closed-form
bound-state energies
 \ben
 E^{(outer)}_\pm=E^{(outer)}_\pm(\gamma)
 =\frac{5}{2}\pm \frac{1}{2}\,\sqrt{17-16\,{{\it \gamma}}^{2}}\,,
 \een
 \ben
 E^{(inner)}_\pm=E^{(inner)}_\pm(\delta)
 =\frac{5}{2}\pm \frac{1}{2}\,\sqrt{1-16\,{{\it \delta}}^{2}}\,.
 \een
We witness the neat separation of the roles of the respective
``amended" coupling constants $\gamma\in
(-\gamma_{(max)},\gamma_{(max)})$ and $\delta\in
(-\delta_{(max)},\delta_{(max)})$ with their respective maxima
compatible with the reality of spectrum being reached at
$\gamma_{(max)}=\pm \sqrt{17/16}$ and $\delta_{(max)}=\pm 1/4$.

%
\begin{figure}[h]                     
\begin{center}                         
\epsfig{file=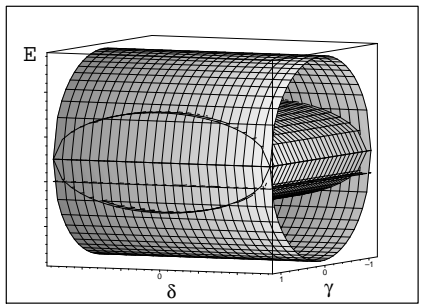,angle=270,width=0.4\textwidth}
\end{center}                         
\vspace{-2mm} \caption{Quadruplet of eigenvalues at $K=1$ and the
separable nature of their complexification pattern.
 \label{fionej2}}
\end{figure}

\subsection{$K=2$: Four-vertex model}

Curiously enough, the above-mentioned separation of the roles of
couplings survives the transition to matrix dimensions $N=2K+2$ with
$K>1$. In particular, the use of couplings $\gamma$ and $\delta$
simplifies the study of the discrete $K=2$ quantum graph
 \be
 \ba
   \mbox{\ \ \ \ \ \ \ \
  }_\diagup\begin{array}{|c|}
 \hline
 {x_{0^+}}\\
 \hline
 \ea_\diagdown \\
 \ \ \ \ \ \
  \begin{array}{||c||}
 \hline
 \hline
 {x_{-2}}\\
 \hline
 \hline
 \ea-
 \begin{array}{||c||}
 \hline
 \hline
 \mbox{} x_{-1}\\
 \hline
 \hline
 \ea_\diagdown^\diagup
 \
 \begin{array}{c}
  \mbox{\ \ \
  \ \ \ \
 \ \ }\\
  \ea
 \ ^\diagdown_\diagup\begin{array}{||c||}
 \hline
 \hline
 \mbox{} x_1\\
 \hline
 \hline
 \ea - \begin{array}{||c||}
 \hline
 \hline
 {x_{2}}\\
 \hline
 \hline
 \ea \,.\\
   \mbox{\ \ \ \ \ \ \ \
 }^\diagdown\begin{array}{|c|}
 \hline
 {x_{0^-}}\\
 \hline
 \ea^\diagup  \\
  \ea
  \ \
 \label{Voo6KL1}
 \ee
Its spectrum is easily shown to be composed of the degenerate
constant doublet $E^{(constant)}_\pm=2$ complemented by the
quadruplet
 \ben
 E^{(outer)}_\pm=E^{(outer)}_\pm(\gamma)
 =\frac{5}{2}\pm \frac{1}{2}\,\sqrt{21-16\,{{\it \gamma}}^{2}}\,,
 \een
 \ben
 E^{(inner)}_\pm=E^{(inner)}_\pm(\delta)
 =\frac{5}{2}\pm \frac{1}{2}\,\sqrt{5-16\,{{\it \delta}}^{2}}\,.
 \een
For nonconstant energies in square-shaped domain of admissible
couplings $\gamma\in (-\gamma_{(max)},\gamma_{(max)})$ and
$\delta\in (-\delta_{(max)},\delta_{(max)})$ we have
$\gamma_{(max)}=\pm \sqrt{21/16}$ and $\delta_{(max)}=\pm
\sqrt{5/16}$.

\subsection{Secular equation at $K=3$ \label{je3}}

The encouraging experience with the above-described $K=1$ and $K=2$
models (where the secular equation has got nicely factorized)
happens to be confirmed at $K=3$ where an analogous analysis of the
one-loop quantum graph
 \be
 \ba
   \mbox{\ \ \ \ \ \ \ \
  }_\diagup\begin{array}{|c|}
 \hline
 {x_{0^+}}\\
 \hline
 \ea_\diagdown \\
 \ \ \
  \begin{array}{||c||}
 \hline
 \hline
 {x_{-3}}\\
 \hline
 \hline
 \ea   -\begin{array}{|c|}
 \hline
 {x_{-2}}\\
 \hline
 \ea-
 \begin{array}{||c||}
 \hline
 \hline
 \mbox{} x_{-1}\\
 \hline
 \hline
 \ea_\diagdown^\diagup
 \
 \begin{array}{c}
  \mbox{\ \ \
  \ \ \ \
 \ \ }\\
  \ea
 \ ^\diagdown_\diagup\begin{array}{||c||}
 \hline
 \hline
 \mbox{} x_1\\
 \hline
 \hline
 \ea - \begin{array}{|c|}
 \hline
 {x_{2}}\\
 \hline
 \ea -
   \begin{array}{||c||}
 \hline
 \hline
 {x_{3}}\\
 \hline
 \hline
 \ea  \\
   \mbox{\ \ \ \ \ \ \ \
 }^\diagdown\begin{array}{|c|}
 \hline
 {x_{0^-}}\\
 \hline
 \ea^\diagup  \\
  \ea
  \
 \label{Vooo8KL2}
 \ee
immediately leads to the secular equation which is factorized in the
same manner as above. This means that the energies are given as
roots of one of the following two polynomial equations
 \be
 {E}^4-9\,{E}^3+
 P_\pm
 \,{E}^2+
 Q_\pm
 \,{E}+
 R_\pm
 =0
 \label{pair}
 \ee
with different respective coefficients
 \ben
 P_+=P_+(z,\gamma)=
 z^2+24+4\,\gamma^2\,,
 \ \ \ \
 Q_+=Q_+(z,\gamma)=-5\,z^2-19-16\,\gamma^2\,,
 \een
 \be
 \ \ \ \
 R_+=R_+(z,\gamma)=
 2\,z^2+4\,\gamma^2\,z^2+12\,\gamma^2+2\,
 \label{pairplus}
 \ee
(which do not depend on $\delta$) and
 \ben
 P_-=P_-(z,\delta)=
 28+z^2+4\,\delta^2\,,
 \ \ \ \
 Q_-=Q_-(z,\delta)=-35-5\,z^2-16\,\delta^2\,,
 \een
 \be
 \ \ \ \
 R_-=R_-(z,\delta)=
 14+6\,z^2+12\,\delta^2+4\,\delta^2\,z^2\,
 \label{pairminus}
 \ee
(which do not depend on $\gamma$). At each $z$ the spectrum is
composed of the two independent one-parametric quadruplets of
levels. The model is easily tractable numerically and its recent
graphical analysis \cite{web} revealed also a number of
phenomenologically interesting features of its energy levels.


\subsection{Secular equation at $K=4$}

%
%
%
%
%

Also the next discrete quantum graph
 \be
 \ba
   \mbox{\ \ \ \ \ \ \ \
  }_\diagup\begin{array}{|c|}
 \hline
 {x_{0^+}}\\
 \hline
 \ea_\diagdown \\
 \
  \begin{array}{||c||}
 \hline
 \hline
 {x_{-4}}\\
 \hline
 \hline
 \ea   -
  \begin{array}{|c|}
 \hline
 {x_{-3}}\\
 \hline
 \ea   -
 \begin{array}{|c|}
 \hline
 {x_{-2}}\\
 \hline
 \ea-
 \begin{array}{||c||}
 \hline
 \hline
 \mbox{} x_{-1}\\
 \hline
 \hline
 \ea_\diagdown^\diagup
 \
 \begin{array}{c}
  \mbox{\ \ \
  \ \ \ \
 \ \ }\\
  \ea
 \ ^\diagdown_\diagup\begin{array}{||c||}
 \hline
 \hline
 \mbox{} x_1\\
 \hline
 \hline
 \ea - \begin{array}{|c|}
 \hline
 {x_{2}}\\
 \hline
 \ea -
   \begin{array}{|c|}
 \hline
 {x_{3}}\\
 \hline
 \ea -
  \begin{array}{||c||}
 \hline
 \hline
 {x_{4}}\\
 \hline
 \hline
 \ea    \\
   \mbox{\ \ \ \ \ \ \ \
 }^\diagdown\begin{array}{|c|}
 \hline
 {x_{0^-}}\\
 \hline
 \ea^\diagup  \\
  \ea
  \
 \label{Vooo10KL3}
 \ee
leads to the eigenvalue problem which is solvable in closed form.
Indeed, the elementary extraction of the two degenerate constant
energies $E^{(constant)}_\pm =2$ leaves the secular equation
factorized into the same two quartic polynomial equations
(\ref{pair}) as above, with just the slightly modified coefficients
 \ben
 P_+=P_+(z,\gamma)=
 z^2+23+4\,\gamma^2\,,
 \ \ \ \
 Q_+=Q_+(z,\gamma)=-5\,z^2-14-16\,\gamma^2\,,
 \een
 \be
 \ \ \ \
 R_+=R_+(z,\gamma)=
 z^2+4\,\gamma^2\,z^2+8\,\gamma^2+1\,
 \label{pairplusbe}
 \ee
 \ben
 P_-=P_-(z,\delta)=
 27+z^2+4\,\delta^2\,,
 \ \ \ \
 Q_-=Q_-(z,\delta)=-30-5\,z^2-16\,\delta^2\,,
 \een
 \be
 \ \ \ \
 R_-=R_-(z,\delta)=
 9+5\,z^2+8\,\delta^2+4\,\delta^2\,z^2\,.
 \label{pairminusbe}
 \ee
We see that even at $K=4$ the factorizability as well as the
survival of the complete separation of the coupling dependence in
the secular equation holds and happens to keep it solvable in closed
form, in principle at least.


\subsection{Secular equation at $K=5$}

The next, $K=5$ version of the discrete quantum graph (\ref{VoooKL})
specifies the next eigenvalue problem which still preserves several
features of its predecessors. Firstly, it gets factorized into two
polynomial subproblems of sixth degree,
 \be
 {E}^6-13\,{E}^5+
 P_\pm
 \,{E}^4+
 Q_\pm
 \,{E}^3+
 R_\pm
 \,{E}^2+
 S_\pm
 \,{E}+
 T_\pm
 =0\,.
 \label{pairbe}
 \ee
Secondly, the individual coefficients still exhibit the same
separation of the couplings as above, having the explicit form
 \ben
 P_+=P_+(z,\gamma)=
 z^2+62+4\,\gamma^2\,,
 \ \ \ \
 Q_+=Q_+(z,\gamma)=-9\,z^2-133-32\,\gamma^2\,,
 \een
 \ben
 \ \ \ \
 R_+=R_+(z,\gamma)=
 24\,z^2+4\,\gamma^2\,z^2+84\,\gamma^2+125\,
 \een
 \ben
 \ \ \ \
 S_+=S_+(z,\gamma)=-19\,z^2-41-80\,\gamma^2\,
 -16\,\gamma^2\,z^2\,
 \een
 \be
 \ \ \ \
 T_+=T_+(z,\gamma)=
 2\,z^2+12\,\gamma^2\,z^2+20\,\gamma^2+2\,
  \label{pairplusbece}
 \ee
and
 \ben
 P_-=P_-(z,\delta)=
 z^2+66+4\,\delta^2\,,
 \ \ \ \
 Q_-=Q_-(z,\delta)=-9\,z^2-165-32\,\delta^2\,,
 \een
 \ben
 \ \ \ \
 R_-=R_-(z,\delta)=
 28\,z^2+4\,\delta^2\,z^2+84\,\delta^2+209\,
 \een
 \ben
 \ \ \ \
 S_-=S_-(z,\delta)=-35\,z^2-121-80\,\delta^2\,
 -16\,\delta^2\,z^2\,
 \een
 \be
 \ \ \ \
 T_-=T_-(z,\delta)=
 14\,z^2+12\,\delta^2\,z^2+20\,\delta^2+22\,.
  \label{pairplusbeceda}
 \ee
This renders  a comfortable numerical analysis possible. Moreover,
as long as even at $K=5$ the factorizability and the separation of
the couplings still holds true, we may reasonably expect its
validity extended to all the integers $K$.


\section{Summary}

The specific coupling-dependence proposed in our present family of
toy-model Hamiltonians has been shown to comply not only with the
preservation of parallels between different models but also with an
enhanced feasibility of numerical experiments with the spectra.
Unexpected regularities were revealed to occur in secular equations,
rendering their complicated three-parametric form still accessible
to non-numerical analysis.

In particular, we confirmed the reality of some of the spectra (and,
hence, the observability of our quantum systems) in fairly large
domains ${\cal D}$ of parameters. Hence, the next task of their
analysis will lie in a constructive replacement of our present
Hamiltonians by their isospectral partners. One can notice that in
the context of quantum theory on graphs, a few simpler, tree-graph
samples of such a replacement have already been non-numerically
constructed in our recent paper~\cite{fundgra}. This makes our
present text tractable as a continuation of the series
\cite{fund,fundgra} in which a synthesis is being sought between
advanced kinematics (topologically nontrivial quantum graphs are
considered) and advanced dynamics (non-Hermitian point-like
interactions are assumed attached to certain graph-shaped
topological anomalies of a restricted, fundamental-length size
${\cal O}(\theta)$).

In this sense, our present models replace a small subinterval of the
real line of coordinates (in its discrete approximation) by a loop.
This introduces a short-range ``smearing of kinematics" which
complements the ``smearing of dynamics" mediated by non-Hermitian
interactions and studied in papers \cite{fund,fundgra}. A wealth of
unusual features may be expected to appear in the spectra of
energies as a consequence \cite{web}. In parallel, many mathematical
difficulties might prove circumvented by the discretization
technique.

In a broader context of quantum model-building effort our present
demonstration of factorizability of secular equations in several
models possessing nontrivial non-Hermitian Hamiltonians is
encouraging. Such an (unexpected) simplification of their algebraic
treatment might inspire their more systematic study in the nearest
future.

\subsection*{Acknowledgements}

The support by the Institutional Research Plan AV0Z10480505 and by
the M\v{S}MT ``Doppler Institute" project LC06002  is acknowledged.

\end{document}